\documentclass[prl,twocolumn,showpacs]{revtex4}
\usepackage{amsmath}
\usepackage{amssymb}
\usepackage{graphicx}
\usepackage{bm}
\usepackage{dcolumn}

\newcommand{\bpm}{\begin{pmatrix}}
\newcommand{\epm}{\end{pmatrix}}

\begin{document}

\title{Slow oscillations of in-plane magnetoresistance in strongly
anisotropic quasi-two-dimensional rare-earth tritellurides}

\author{P.D.~Grigoriev$^{1}$, A.A.~Sinchenko$^{2,3,4,5}$, P.~Lejay$^{3,4}$, O. Leynaud$^{3,4}$, V.N.~Zverev$^{6,7}$ and P.~Monceau$^{3,4}$}

\address{$^{1}$L. D. Landau Institute for Theoretical Physics, 142432
Chernogolovka, Russia}

\address{$^{2}$Kotel'nikov Institute of Radioengineering and Electronics of RAS, Mokhovaya 11-7, 125009 Moscow, Russia}

\address{$^{3}$Univ. Grenoble Alpes, Inst. Neel, F-38042 Grenoble, France}

\address{$^{4}$CNRS, Inst. Neel, F-38042 Grenoble, France}

\address{$^{5}$National Research Nuclear University (MEPhI), 115409 Moscow,Russia}

\address{$^{6}$Institute of Solid State Physics, Chernogolovka, Moscow region 142432, Russia}

\address{$^{7}$Moscow Institute of Physics and Technology, Dolgoprudnyi, Moscow region, 141700, Russia}

\date{\today }

\begin{abstract}
Slow oscillations of the in-plane magnetoresistance are observed
in the rare-earth tritellurides and proposed as an effective tool
to determine the parameters of electronic structure in various
strongly anisotropic quasi-two-dimensional compounds. These
oscillations do not originate from the small Fermi surface
pockets, as revealed usually by the Shubnikov-de-Haas
oscillations, but from the entanglement of close frequencies due
to a finite interlayer transfer integral $t_z$, which allows to
estimate its value. For TbTe$_3$ and GdTe$_3$ we obtain the
estimate $t_z\approx 1$ meV.
\end{abstract}


\maketitle

The angular and magnetic quantum oscillations (MQO) of
magnetoresistance (MR) is a powerful tool to study electronic
properties of various quasi-two-dimensional (Q2D) layered metallic
compounds, such as organic metals (see, e.g., Refs.
\cite{MQORev,OMRev,MarkReview2004,KartPeschReview} for reviews),
cuprate and iron-based high-temperature superconductors (see,
e.g.,
\cite{HusseyNature2003,AbdelNature2006,ProustNature2007,AbdelPRL2007AMRO,McKenzie2007,DVignolle2008,HelmNd2009,HelmNd2010,BaFeAs2011,Graf2012}),
heterostructures\cite{Kuraguchi2003} etc.

The Fermi surface (FS) of Q2D metals is a cylinder with weak
warping $\sim 4t_{z}/E_{F}\ll 1$, where $t_{z}$ is the interlayer
transfer integral and $E_{F}=\mu $ is the in-plane Fermi energy.
The MQO with such FS have two close fundamental frequencies
$F_{0}\pm \Delta F$. In a magnetic field $\boldsymbol{B}=B_{z}$
perpendicular to the conducting layers $F_{0}/B=\mu /\hbar \omega
_{c}$ and $\Delta F/B=2t_{z}/\hbar \omega _{c} $, where  $\hbar
\omega _{c}=\hbar eB_{z}/m^{\ast }c$ is the separation between the
Landau levels (LL).

The standard 3D theory of galvanomagnetic properties
\cite{Abrik,Shoenberg,Ziman} is valid only at $t_{z}\gg \hbar
\omega _{c}$ and in the lowest order in the parameter $\hbar
\omega _{c}/t_{z}$. This theory predicts several peculiarities of
MR in Q2D metals: the angular magnetoresistance oscillations
(AMRO)\cite{KartsAMRO1988,Yam,Yagi1990} and the beats of the
amplitude of MQO \cite{Shoenberg}. One can even extract the fine
details of the FS, such as its in-plane anisotropy \cite{Mark92}
and its harmonic expansion,\cite{Bergemann,GrigAMRO2010} from the
angular dependence of MQO frequencies and from AMRO. For isotropic
in-plane electron dispersion, AMRO can be described by the
renormalization of the interlayer transfer integral:\cite{Kur}
\begin{equation}
t_{z}=t_{z}\left( \theta \right) =t_{z}\left( 0\right) J_{0}\left(
k_{F}d\tan \theta \right) ,  \label{tz}
\end{equation}%
where $J_{0}\left( x\right) $ is the Bessel's function,
$p_{F}=\hbar k_{F}$ is the in-plane Fermi momentum, and $\theta $
is the angle between magnetic field $\boldsymbol{B}$ and the
normal to the conducting layers.

At $t_{z}\sim \hbar \omega _{c}$ new qualitative features appear
both in the monotonic and oscillating parts of MR. For example,
the strong monotonic growth of interlayer MR $R_{zz} (B_{z})$ was
observed in various Q2D metals
\cite{SO,Coldea,PesotskiiJETP95,Zuo1999,W3,W4,Incoh2009,Wosnitza2002,Kang,WIPRB2012}
and recently theoretically explained
\cite{WIPRB2012,WIPRB2011,WIPRB2013,GG2014}. The oscillating part
of interlayer MR at $\mu \gg t_{z}\gtrsim \hbar \omega _{c}$
acquires the slow oscillations \cite{SO,Shub} and the phase shift
of beats.\cite{PhSh,Shub} These two effects are missed by the
standard 3D theory \cite{Abrik,Shoenberg,Ziman} because they
appear in the higher orders in $\hbar \omega _{c}/t_{z}$.

The slow oscillations qualitatively originate from the product of
the oscillations with two close frequencies $F_{0}\pm \Delta F$,
which gives the oscillations with frequency $2\Delta F$.
Conductivity, being a non-linear function of the oscillating
electronic density of states (DoS) and of the diffusion
coefficient, has slow oscillations with frequency $2\,\Delta F$,
while magnetization, being a linear functional of DoS, does not
show slow oscillations \cite{SO,Shub}. The slow oscillations have
many interesting and useful features as compared to the fast
quantum oscillations. First, they survive at much higher
temperature than MQO. Second, they are not sensitive to the
long-range disorder, which has a strong action on fast MQO,
similar to the influence of finite temperature. Third, the slow
oscillations allow to measure the interlayer transfer integral
$t_{z}$. These features make the slow oscillations to be a useful
tool to study the electronic properties of Q2D metals. Until now,
the slow oscillations where investigated only for the interlayer
conductivity $\sigma _{zz}\left( B\right) $, when the current and
the magnetic field are both applied perpendicularly to the 2D
layers, and only in the organic compounds. In this paper, we study
the slow oscillations of transverse intralayer MR
$R_{yy}(\boldsymbol{B})$ both theoretically and experimentally in
the non-organic rare-earth tritelluride Q2D layered compounds.

The rare-earth tritelluride compounds $R$Te$_3$ ($R=$Y, La, Ce, Nd,
Sm, Gd, Tb, Ho, Dy, Er, Tm) have the same orthorhombic structure
($Cmcm$) in the normal state. These systems exhibit a $c$-axis
incommensurate charge-density wave (CDW) at high temperature through
the whole $R$ series that was recently a subject of intense studies
\cite{DiMasi95,Brouet08,Ru08,SinchPRB12,Anis13,SSC14}. For the
heaviest rare-earth elements, a second $a$-axis CDW occurs at low
temperature. In addition to hosting incommensurate CDW, magnetic
rare-earth ions exhibit closed-spaced magnetic phase transitions
below 10 K \cite{Iyeri2003, Ru2008} leading to coexistence and
competition of many ordered states at low temperature. Therefore,
any information about the Fermi surface warping in the Q2D-compounds
(determined by $t_z$) is very important. For the possible
observation of the slow oscillations the rare-earth tritellurides
are very promising, because they have appropriate anisotropy and
good metallic conductivity up to low temperatures.

For experiments we choose GdTe$_3$ and TbTe$_3$. Single crystals of
these compounds were grown by a self-flux technique under purified
argon atmosphere as described previously \cite{SinchPRB12}. Thin
single crystal samples with a thickness typically 0.1-0.3 $\mu$m
were prepared by micromechanical exfoliation of relatively thick
crystals glued on a sapphire substrate. The quality of selected
crystals and the spatial arrangement of crystallographic axes were
controlled by the X-ray diffraction. From high-quality $[R(300
K)/R(10 K)>100]$ untwinned single crystals we cut bridges with a
width $50-80$ $\mu$m in well defined, namely [100] and [001],
orientations. Contacts for electrical transport measurements in the
four-probe configuration have been prepared using gold evaporation
and cold soldering by In. Magnetotransport measurements were
performed at different orientations of the magnetic field in the
field range up to 9 T using a superconducting solenoid. The field
orientation was defined by the angle $\theta $ between the field
direction and the normal to the highly conducting $ac$ plane.

\begin{figure}[t]
\includegraphics[width=8.5cm]{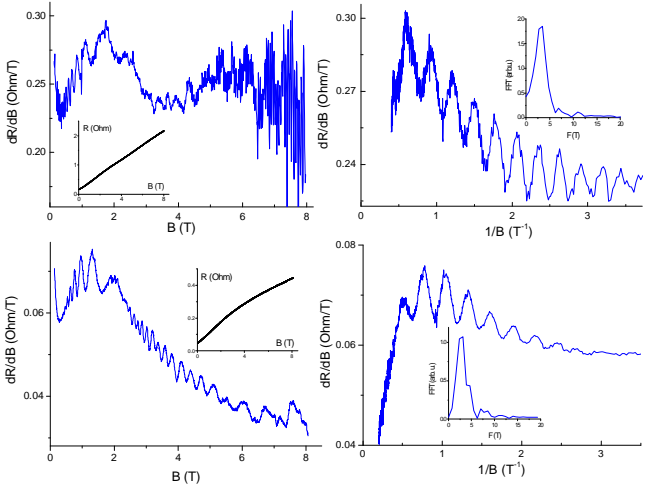}
\caption{(color online) $dR(B)/dB$ dependencies at $T=4.2$ K for
TbTe$_{3}$ (a) and GdTe$_{3}$ (c) demonstrating clear rapid
oscillations which appear in the field range $B>2$ T. Insets show
magnetoresistance $R(B)$. (b) and (d) -- $dR/dB$ as a function of
inverse magnetic field, $B^{-1}$, at low field region
demonstrating slow oscillations in TbTe$_{3}$ and GdTe$_{3}$
correspondingly. Insets show the Fourier transform of slow
oscillations.} \label{F1}
\end{figure}

In Figure \ref{F1} we plot the derivative of MR $dR/dB$ as a
function of magnetic field up to $B=8.2$ T applied along $b$-axis
for TbTe$_{3}$ (a) and GdTe$_{3}$ (c) with the current applied in
the $(a,c)$ plane at $T=4.2$ K. For both measured compounds, at
$B>2$T the pronounced Shubnikov - de Haas (SdH) oscillations with
a frequency $F\approx 55-58$ T are observed in the derivative of
MR, $dR/dB$. At high field ($B\gtrsim 7$ T) new oscillations with
higher frequency ($F\approx 0.7-0.8$ kT) appear, indicating the
existence of several types of pockets on the partially gapped FS,
resulting from the CDW formation \cite{LaTe08}. Insets in Fig.
\ref{F1}a,c show the corresponding MR curves indicating very weak
amplitude of oscillations. In addition to rapid SdH oscillations,
at low magnetic field ($B<2$ T) the MR exhibits prominent slow
oscillations with very low frequency $F_{slow}\lesssim 4$ T. To
demonstrate the periodicity of these oscillations in $B^{-1}$, we
plot the derivative $dR/dB$ as a function of inverse magnetic
field in Fig. \ref{F1} for TbTe$_{3}$ (b) and for GdTe$_{3}$ (d).
Insets show the Fourier transform of $dR/dB(1/B)$.

\begin{figure}[t]
\includegraphics[width=8cm]{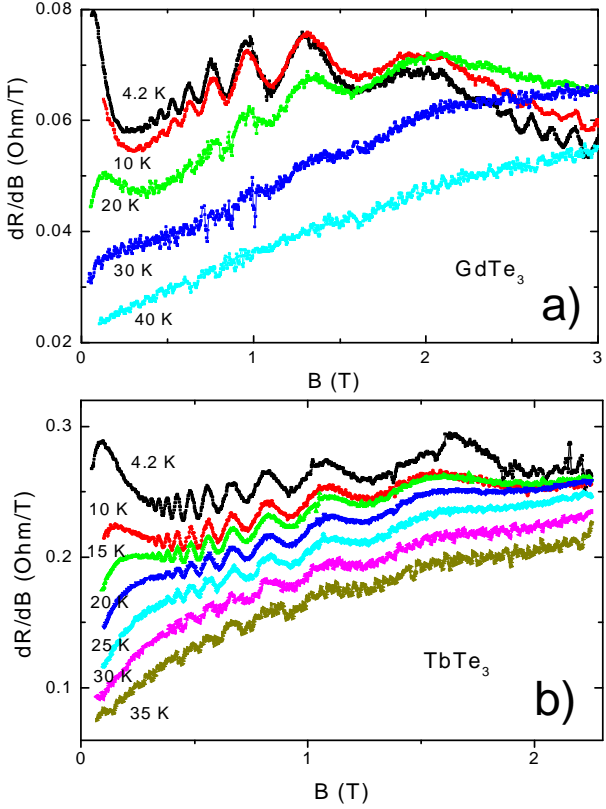}
\caption{(color online) Temperature evolution of slow oscillations
for GdTe$_3$ (a) and TbTe$_3$ (b).} \label{F2}
\end{figure}

In contrast to the usual SdH oscillations, the amplitude of which
decreases rapidly as temperature increases, the slow oscillations
of MR are observable up to $T\backsimeq 40$ K, as can be seen from
Fig. \ref{F2} where we show the temperature evolution of slow
oscillations for GdTe$_{3}$ and TbTe$_{3}$. This suggests that the
observed slow oscillations originate not from small FS pockets,
but from the FS warping due to $t_{z}$, similarly to the slow
oscillations of interlayer MR in the organic superconductor $\beta
$-(BEDT-TTF)$_{2}$IBr$_{2}$ \cite{SO}. If so, the observed slow
oscillations give an excellent opportunity to find the value of
$t_{z}$ and of the product $k_{F}d$ at low temperature in
tritellurides TbTe$_{3}$ and GdTe$_{3}$ from the experimental data
on the intralayer MR.

According to Eq. (90.5) of Ref. \cite{LL10} the intralayer
conductivity at finite temperature is given by \cite{LL10}
\begin{equation}
\sigma _{yy}=\int d\varepsilon \,\left[ -n_{F}^{\prime }(\varepsilon
)\right] \,\sigma _{yy}(\varepsilon ),  \label{sigmayy}
\end{equation}%
where the derivative of the Fermi distribution function
$n_{F}^{\prime }(\varepsilon )=-1/\{4T\cosh ^{2}\left[
(\varepsilon -\mu )/2T\right] \}$, and the zero-temperature
electron conductivity at energy $\varepsilon $ is
\begin{equation}
\sigma _{yy}(\varepsilon )=e^{2}g\left( \varepsilon \right)
D_{y}\left( \varepsilon \right) .  \label{s1}
\end{equation}%
Here $g\left( \varepsilon \right) $ is the DoS and $D_{y}\left(
\varepsilon \right) $ is the diffusion coefficient of electrons along
y-axis. It is convenient to use the harmonic expansion for the
oscillating DoS $g\left( \varepsilon \right) $. Below we will need
only the first terms in this harmonic series, which at finite
$t_{z}\sim \hbar \omega _{c}$ are given by \cite{Champel2001,ChampelMineev,Shub}%
\begin{equation}
g\left( \varepsilon \right) \approx g_{0}\left[ 1-2\cos \left(
\frac{2\pi \varepsilon }{\hbar \omega _{c}}\right) J_{0}\left(
\frac{4\pi t_{z}}{\hbar \omega _{c}}\right) R_{D}\right] ,
\label{ge1}
\end{equation}
where $g_{0}=m^{\ast }/\pi \hbar ^{2}d$ is the DoS at the Fermi
level in the absence of magnetic field per two spin components,
$J_{0}\left( x\right) $ is the Bessel's function, the Dingle
factor\cite{Dingle,Bychkov} $R_{D}\approx \exp \left[ -\pi k/\omega _{c}\tau
_{0}\right] $, $\tau _{0}=\hbar /2\Gamma _{0}$ is the electron
mean free time without magnetic field, and $\Gamma _{0}$ is the LL
broadening.

The calculation of the diffusion coefficient $D_{y}\left(
\varepsilon \right) $ is less trivial and requires to specify the
model. At $\mu \gg \hbar \omega _{c}$ the quasi-classical
approximation is applicable. In an ideal crystal in a magnetic
field $\boldsymbol{B}$ the electrons move along the cyclotron
orbits with a fixed center and the Larmor radius $
R_{L}=p_{F}c/eB_{z}$. Without scattering the electron diffusion in
the direction perpendicular to $\boldsymbol{B}$ is absent.
Scattering by impurities changes the electronic states and leads
to the electron diffusion. For simplicity, we consider only
short-range impurities, described by the $\delta $-function
potential: $V_{i}\left( r\right) =U\delta ^{3}\left(
r-r_{i}\right) $. Scattering by impurities is elastic, i.e. it
conserves the electron energy $\varepsilon $, but the quantum
numbers of electron states may change. The matrix element of
impurity scattering is given by
\begin{equation}
T_{mm^{\prime }}=\Psi _{m^{\prime }}^{\ast }\left( r_{i}\right)
U\Psi _{m}\left( r_{i}\right) ,  \label{Tm}
\end{equation}%
where $\Psi _{m}\left( r\right) $ is the electron wave function in
the state $m$. During each scattering, the typical change $\Delta
y=\Delta P_{x}c/eB_{z}$ of the mean electron coordinate $y_{0}$
perpendicular to $\boldsymbol{B}$ is of the order of $R_{L}$,
because for larger $\Delta y\gg R_{L}$ the matrix element in Eq.
(\ref{Tm}) is exponentially small because of small overlap of the
electron wave functions $\Psi _{m^{\prime }}^{\ast }\left(
r_{i}\right) \Psi _{m}\left( r_{i}\right) \sim \Psi _{m}^{\ast
}\left( r_{i}+\Delta y\right) \Psi _{m}\left( r_{i}\right) $
\cite{CommentDecay}. The diffusion coefficient is approximately
given by
\begin{equation}
D_{y}\left( \varepsilon \right) \approx \left\langle \left( \Delta
y\right) ^{2}\right\rangle /2\tau \left( \varepsilon \right) ,
\label{D1}
\end{equation}%
where $\tau \left( \varepsilon \right) $ is the energy-dependent electron mean scattering time
by impurities, and the angular brackets in Eq. (\ref{D1}) mean
averaging over impurity scattering events. In the Born
approximation, the mean scattering rate
\begin{equation}
1/\tau \left( \varepsilon \right) =2\pi n_{i}U^{2}g\left(
\varepsilon \right) ,  \label{tau}
\end{equation}%
where $n_{i}$ is the impurity concentration. This scattering rate
has MQO, proportional to those of the DoS in Eq. (\ref{ge1}). The
MQO of $ \left\langle \left( \Delta y\right) ^{2}\right\rangle
\approx R_{L}^{2}  $ are, usually, weaker and in 3D metals they are
neglected \cite{LL10}. Then
\begin{equation}
D_{y}\left( \varepsilon \right) \approx R_{L}^{2}/2\tau \left(
\varepsilon \right) \propto g\left( \varepsilon \right) . \label{D2}
\end{equation}%
\ However, in Q2D metals, when $t_{z}\sim \hbar \omega _{c}$, the
MQO of $\left\langle \left( \Delta y\right) ^{2}\right\rangle $ can
be of the same order as MQO of the DoS. Then instead of Eq.
(\ref{D2}) at $R_{D}\ll 1$ one has
\begin{equation}
D_{y}\left( \varepsilon \right) \approx D_{0}\left[ 1-2\alpha \cos
\left( \frac{2\pi \varepsilon }{\hbar \omega _{c}}\right)
J_{0}\left( \frac{4\pi t_{z}}{\hbar \omega _{c}}\right) R_{D}\right]
,  \label{D3}
\end{equation}%
where $D_{0}\approx R_{L}^{2}/2\tau _{0}$, and the number $\alpha
\sim 1$. Substituting Eqs. (\ref{s1}), (\ref{ge1}) and
(\ref{D3}) to Eq. (\ref{sigmayy}) after integration over $\varepsilon $ one obtains%
\begin{gather}
\frac{\sigma _{yy}(B)}{e^{2}g_{0}D_{0}}\approx 1+2\alpha J_{0}^{2}\left(4\pi t_{z}/\hbar \omega _{c}\right) R_{D}^{2}-\label{SO} \\
-2\left( \alpha +1\right) \cos \left( \frac{2\pi \mu }{\hbar \omega
_{c}}\right) J_{0}\left( \frac{4\pi t_{z}}{\hbar \omega _{c}}\right)
R_{D}R_{T}, \notag
\end{gather}
where the temperature damping factor of MQO is
\begin{equation}
R_{T}=\left( 2\pi ^{2}k_{B}T/\hbar \omega _{c}\right) /\sinh \left(
2\pi ^{2}k_{B}T/\hbar \omega _{c}\right) .  \label{RT}
\end{equation}%
The slow oscillations, described by the first line of Eq.
(\ref{SO}), are not damped by temperature within our model.

Approximately, one can use the asymptotic expansion of the Bessel
function in Eq. (\ref{SO}) for large values of the argument:
$J_{0}(x)\approx \sqrt{2/\pi x}\cos \left(x-\pi/4\right)\,,\,x\gg
1$. Then, after introducing the frequency of slow oscillations
$F_{slow}= 4t_{z}B/\hbar \omega _{c}$, the first line in Eq.
(\ref{SO}) simplifies to
\begin{equation}
\frac{\sigma ^{slow}_{yy}(B)}{e^{2}g_{0}D_{0}}\approx
1+\frac{\alpha \hbar \omega _{c}}{2\pi ^{2}t_{z}}\sin \left(
\frac{2\pi F_{slow}}{B}\right) R_{D}^{2}. \label{SOa}
\end{equation}%

In tilted magnetic field at constant $\left\vert
\boldsymbol{B}\right\vert $, $\omega _{c}\propto \cos\theta$ and
$t_{z}$ changes according to Eq. (\ref{tz}). Then the frequency of
slow oscillations must depend on tilt angle $\theta $ as:
\begin{equation}
F_{slow}\left( \theta \right) /F_{slow}\left( 0\right)
=J_{0}\left( k_{F}d\tan \theta \right) /\cos \left( \theta
\right) . \label{Fslow}
\end{equation}

\begin{figure}[t]
\includegraphics[width=8cm]{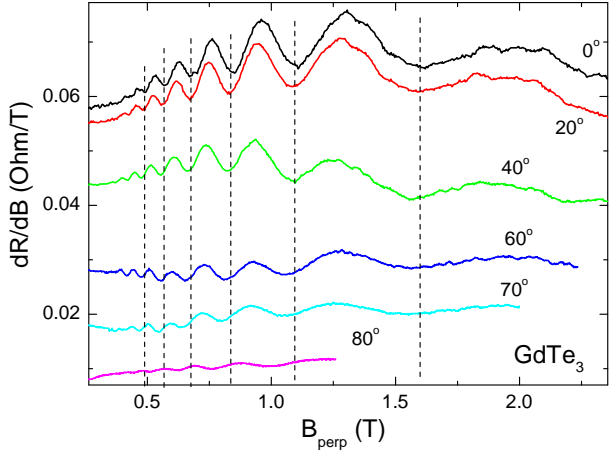}
\caption{(color online) Slow oscillations in GdTe$_{3}$ at $T=4.2$ K for different tilt angles
$\protect\theta $. $B_{\perp }=B\cos (\protect\theta )$.} \label{F3}
\end{figure}

To clarify this effect, we experimentally study the angular
dependence of observed slow oscillations frequency. The evolution of
the slow oscillations in GdTe$_{3}$ with the change of the tilt
angle $\theta $ of magnetic field at $T=4.2$ K is shown in Fig.
\ref{F3}, where the derivative $dR/dB$ is plotted as function of the
perpendicular-to-layers component of magnetic field $B_{\perp
}=B\cos (\theta )$. Note, that magnetic field rotation in the
($b$-$c$) and ($b$-$a$) planes demonstrated the same results for
TbTe$_3$.

\begin{figure}[t]
\includegraphics[width=8.5cm]{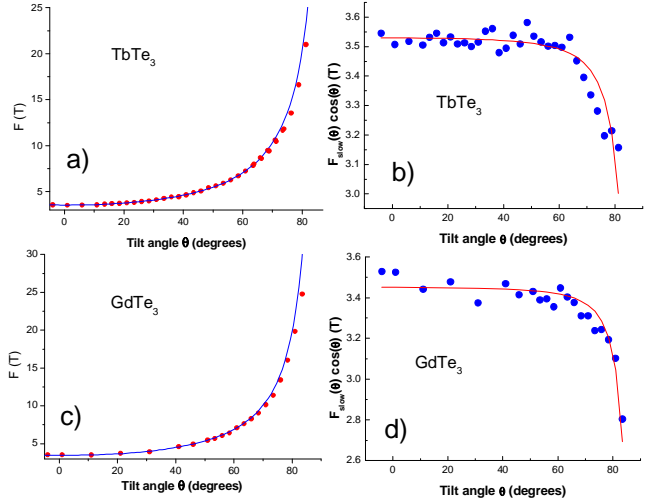}
\caption{(color online) (a) and (c): The frequency of
slow oscillations $F_{slow}$ as a function of tilt angle $\protect\theta $
at $T=4.2$K for TbTe$_{3}$ and GdTe$_{3}$. Solid
curves show the function $F(\protect\theta)=F(0)/\cos (\protect\theta )$.
(b) and (d): Angular dependence of the frequency
$F_{slow}(\protect\theta )$ of slow oscillations of intralayer
magnetoresistance in TbTe$_{3}$ and GdTe$_3$ correspondingly,
multiplied by $\cos (\protect\theta )$. The experimental data are
shown by black filled squares, and the theoretical prediction
according to Eq. (\protect\ref{Fslow}) at $k_{F}d=0.12$ is shown by
solid red lines.} \label{F4}
\end{figure}

In Fig. \ref{F4} we show the $\theta $-dependence of the frequency
of slow oscillations at $T=4.2$ K for TbTe$_{3}$ (a) and for
GdTe$_{3}$ (c). The solid curves give the cosine dependence
$F(\theta )=F(0)/\cos (\theta )$ typical for MQO. According to Eq.
(\ref{Fslow}), the angular dependence of the frequency
$F_{slow}\left( \theta \right) $ of slow oscillations differs from
this standard cosine dependence, especially at high tilt angle. In
Fig. \ref{F4} (b) we plot the angular dependence of the product
$F_{slow}(\theta )\cos (\theta )$ in TbTe$_{3}$. If the origin of
slow oscillations is due to small FS pockets, the product
$F_{slow}(\theta )\cos (\theta )$ would be independent of the tilt
angle $\theta $. The experimental data, shown by black filled
squares, clearly indicate the deviation from the horizontal line.
These experimental data can be reasonably fitted by Eq.
(\ref{Fslow}) at $k_{F}d=0.1$, shown by solid red lines in Figs.
\ref{F4} (b,d).

From the angular dependence of the frequency $F_{slow}\left(
\theta \right)$ of slow oscillations, shown in Fig. \ref{F4}, we
conclude that these oscillations originate not from small FS
pockets as usual SdH oscillations, but from the entanglement of
close frequencies due to a finite $t_{z}$, similar to that
observed in the interlayer MR in the organic metal $\beta
$-(BEDT-TTF)$_{2}$IBr$_{2}$ \cite{SO}. The strong additional
argument in favor of this origin of the observed slow oscillations
is the very weak temperature dependence of their amplitude. To our
knowledge, the data obtained are the first observation of such
slow oscillations in the intralayer magnetotransport.

The frequency of slow oscillations at $\theta =0$ can be used to
estimate the value of $t_{z}$. According to Eq. (\ref{SOa}), with
the effective electron mass $m^{\ast }\approx 0.1m_{e}$ determined
from the temperature dependence of the amplitude of SdH
oscillations \cite{JLTP15}, and $F_{slow}\approx 3.5$T (see Fig.
\ref{F4} (b,d)), we obtain $t_{z}\approx 1meV$. The small values
of $t_z$ compared to the transfer integrals $t_{\parallel}\approx
2$ eV along the chains and $t_{\perp}\approx 0.37$ eV
perpendicular to the chains, as obtained by the band structure
calculations \cite{Brouet08}, illustrate the quasi-2D character of
these rare-earth tritellurides and justify that the dispersion
along $b$-axis is neglected in ARPES measurements. The value of
$t_{z}$ is very important for quantum corrections to conductivity
\cite{AABook,Kennett2008} and for other physical properties of
strongly anisotropic compounds.

The angular dependence of the frequency $F_{slow}\left( \theta
\right) $ of slow oscillations allows also to determine the value
of the Fermi momentum of the open FS pockets. Fitting experimental
data on $F_{slow}\left(\theta \right)$ shown in Fig. \ref{F4} to
Eq. (\ref{Fslow}) gives $k_{F}d\approx 0.11$ for GdTe$_{3}$ and
$k_{F}d\approx 0.12$ for TbTe$_{3}$. With $d\approx 25 $\AA \ this
gives $k_{F}\approx 4.5\cdot 10^5cm^{-1}$. Such a small value of
$k_{F}$ means that the FS reconstruction due to CDW is very strong
and leaves only very small ungapped FS pockets. These FS pockets
are most probably elongated and directed differently, which gives
different $F_{slow}\left(\theta \right)$ dependence. Their total
contribution to the slow oscillations, being a sum of the
contributions from the individual FS pockets, has a smeared
angular dependence of the frequency of slow oscillations
$F_{slow}\left(\theta \right)$ as compared to the case of only one
elliptical FS pocket observed in $\beta
$-(BEDT-TTF)$_{2}$IBr$_{2}$ \cite{SO}.

To summarize, we report the first observation and qualitative
theoretical description of the slow oscillations of intralayer
magnetoresistance in quasi-2D metallic compounds. These slow
oscillations allow to measure the interlayer transfer integral
$t_z$, which is hard to measure by any other ways. We obtain the
value $t_{z}\approx 1meV$ in the rare-earth tritelluride compounds
TbTe$_{3}$ and GdTe$_{3}$.

\acknowledgements

The work was supported partially by RFBR (grants No. 11-02-01379-a
and 13-02-00178-a) and partially performed in the CNRS-RAS
Associated International Laboratory between CRTBT and IRE "Physical
properties of coherent electronic states in coherent matter".

\end{document}